# Dynamical model of spreading of COVID-19 based on a flow networks and SIR family: A comparison of trends in Spain and the Netherlands

*Victoria Lopez, PhD and Milena Čukić, PhD*

Institute of Knowledge Technology, Universidad Complutense de Madrid, Spain

3EGA, Amsterdam, The Netherlands

vlopezlo@ucm.es; micukic@ucm.es

## Abstract

The pandemic of SARS-CoV-2 made many countries impose restrictions in order to control its dangerous effect on the citizens. This work is a theoretical dynamical model based on flow networks and the SIR family model that illustrates the developments and trends based on publicly available data. Based on this model a code in R was written and fed by publicly available data from Spain and Netherlands, to compare their trends. Our results show that the 'peak' of infection is already behind us (in both countries), but also demonstrate that there is a danger of rebound of a spread of infections. It is obvious that measures imposed are giving the results, but we should be precarious of near future practices and development, since the majority of population will be still susceptible to infection. Most importantly, this model can be easily adapted to other regions to replicate model in different situations and be useful for optimization of response.

## 1. Introduction

The viral infection is raging around the world. Since its start in China in last quarter of 2019, the severe acute respiratory syndrome Corona-virus 2 (SARS-CoV-2) which is causing the disease (COVID-19) spread all over the world in last three months (WHO). The first European country registering the person with COVID-19 was France, after which other countries followed. Among them, with very high speed of spread was Spain which already reached the 'peak' of number of infected citizens. In the Netherlands the alarming level of infections was lagging after Spain; the first patient (infected during his recent travel to Lombardy) in the Netherlands was confirmed on February 27th 2020. Very soon after outbreak the website dashboard of John Hopkins University started updating publicly available data on number of infected, deceased and hospitalized persons in all affected countries and many scientific groups tried to employ their knowledge and already developed methods to answer various unknowns in connection to this epidemiologic phenomenon. Many research groups are working on similar problems and we wanted to stress here some of the most important findings so far. Friston published technical paper present on government website in UK (Friston et al., 2020) the Dynamical Causal Model of spread, shedding light on some parts of the mechanism. The same group published another paper (also as a preprint on ArXiv) aiming at estimating the susceptible part of population coming to the conclusion that less than 20% of EU countries population is infected, and that only 6.4% of those infected are gaining immunity. Since their posteriori estimate is suggesting that the majority of European population will stay without the immunity after this first outbreak, this would eventually lead to another cycle of infections, resulting in consecutive peaks. They compared this situation with Spanish Flu pandemic from 2018, which had three peaks in total (Moran et al., 2020). Their result is important since it can inform further governmental actions to prevent loss of larger parts of European population due to the high infectivity of the Corona-virus. Researchers from the Netherlands concentrated on testing of healthcare workers who became infected, concluding that more than half of them had mild symptoms and many of them kept on working while symptomatic (Kluytmans van den Bergh et al., 2020). On RIVM governmental site (The Netherlands, April 10.2020.), it was stated that at one-point healthcare professionals were actually 24% of all recorded infected cases. Chinese researchers reported before about the mechanisms and levels of infectivity, as well as symptoms, but also used the epidemiological tools to predict outcomes (Prem, Liu et al. 2020). Several studies tried to forecast the future disease trajectories informing further management of hospitalization (Moghadas, Shoukat et al., 2020). As we were working on this model the number of preprints regarding COVID-19 doubled every week, resulting in another phenomenon described by the Nature (from 29. April): the publishers removing firewalls and speeding up the reviewing process for COVID-19 bundles actually contributed to slipping in some incompetent or bad science using this opportunity for fast publication.

Although different countries adopted different approaches to solving the problem of massive infections, mainly by social distancing and lockdown, it can be seen that a similar pattern are present. Modelling a dynamic system that represent those patterns has been the goal of the present work. We started with the epidemiological compartments model originated from work of Kermack and McKendrick in 1927 (SIR) and then we developed a flow network with some characteristics from Markov chain where all the states for a person are represented.

The first results of this work were publicly shared on LinkedIn pages of the authors (in consecutive days from the end of March and the first three weeks of April 2020), gaining valuable suggestions from colleagues on how to improve the model. After defining parameters and inferring unknown variables, the code in R was written and fed first by the publicly available data present on John Hopkins University Website (Dong et al., 2019). After the first week, we started using publicly available data present on the website of Spanish Ministry of Health (INE), after which we decided to start calculating the trends for the Netherlands (from governmental RIVM website) also with the difference in the data accessible. For example, the data about the numbers of recovered persons is not stated in RIVM daily updates, it has to be inferred, although the disclaimer says that the number of deaths and hospitalization very often are cumulative for couple of days, especially around Easter holiday. The aim of this work is to compare trends in two European countries since the initial dynamics in one country can become similar in another one, which is lagging in numbers of hospitalized and infected people. It can already be seen that measures taken are affecting the dynamics and total numbers of affected people in both countries. Also, the aim of our work is to share the detailed theoretical model (especially how we determined transfer parameters between states) in order to help further balancing the future model and managing the scarce resources, since we believe that another peak in infection is possible.

This article is organized as follows. After this brief introduction, in Section 2 methods and data are explained. Section 3 is dedicated to the data available for the study and the necessity of inferences or data estimation. We explain all details of our model in Section 4, where it is introduced as a Flow Network Model. Results of the application of the model can be read in section 5. Finally, Section 6 contains the discussion, conclusion and future work.

## 2. Methods and previous models

Compartmental model is technique which simplify the mathematical modelling of infectious disease. One of the simplest compartment models is SIR model. It presumes that all the members of population are going through the three states/belong to three compartments: those who are susceptible (S), those who are infected (I), and eventually those who recovered (R-number of recovered or deceased). To represent that the number of susceptible, infected and recovered individuals may vary over time (even if the total population size remains constant), we make the precise numbers a function of time (t): S(t), I(t) and R(t). The model utilizes ordinary differential equations (ODE) which are deterministic, but the dynamics of the flow can become nonlinear, and therefore can be understood in stochastic framework. If the latter is the case the model becomes more realistic, but also much more complicated to numerically execute.

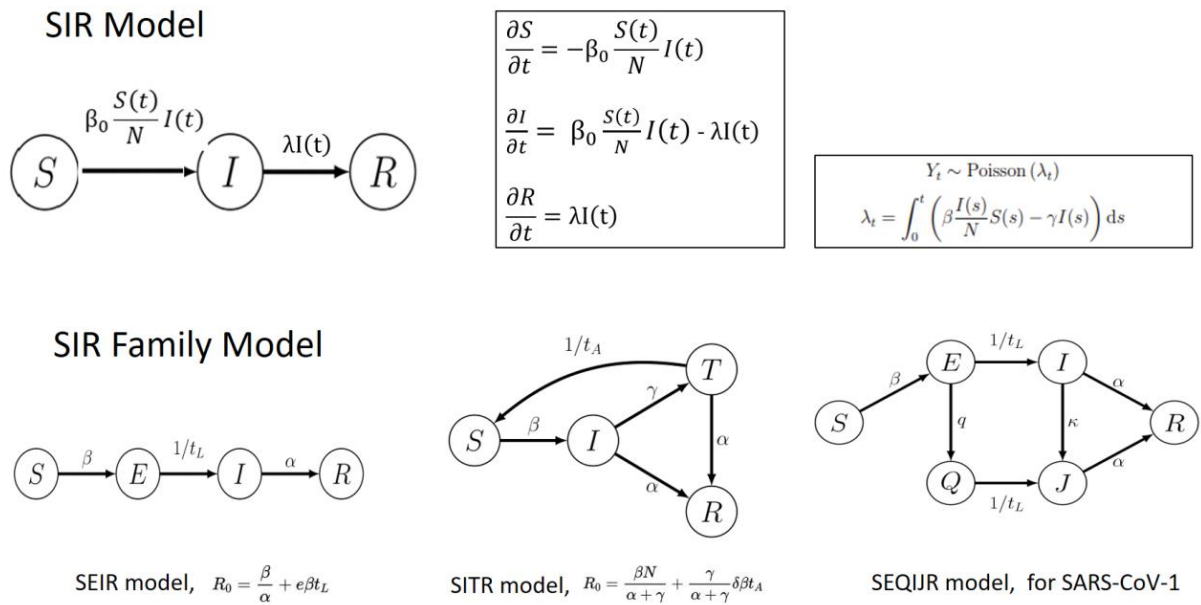

**Figure 1**. SIR Model and SIR Family Model as they were developed by researchers in last decades to deal with similar sanitary crisis.

The SIR model (Susceptible, Infected, Recovered) is deductive. Within the differential equation system t (time unit) is the variable and coefficients are a list of parameters. These parameters are not easy to infer because they are fuzzy, noisy and dependent of the same data from which the model is due. Figure 1 shows the sagittal graph and some details on the different perspectives (SIR. SEIR, SITR, SEQIJR). Also in this figure, the distribution of the data (Poisson) and the calculation of its parameter ($\lambda$) are remarked and the system of derivative equations that models the SIR system: from the definition of the states on day t+1 with respect to the previous day t, and using the derivative approximation, the system of differential equations is constructed. This idea simplified all the others in the family. The SEQIJR model was used by the researchers for SAR-COV-1, the predecessor of SAR-COV-2 (which is causing of COVID-19). These models include nodes for: Susceptible (S), Latent (E), Infected (I), Quarantine (T, Q) Isolation (J) and Recovered (R). The model of our work has 10 nodes and the Quarantine and Isolation nodes are implemented within real containers in the process of recovering as next section shows. The advantage of our model is that the parameters are less uncertain for calculations, as it is explained in Section 4.

## 3. Data

First of all, we started from the information about the number of inhabitants of Spain as a publicly available number of 47 100 396 [www.ine.es] of which 17% are people older than 60 years (of whom 25% are over 80). For the Nederland (RIVM), the National Statistics Office stated that the number of population (the data are from November 2019) is17 424 978 [Centraal Bureau voor de Statistiek, CBS], and that 14.9% of them are persons older than 65.

During the first week we used data from a repository for the 2019 Novel Coronavirus at John's Hopkins University Center for Systems Science and Engineering (JHU CSSE) were used (Dong, Du et al. 2020). After that initial week, we started using instead the official reports of Spanish (www.ine.es, https://covid19.isciii.es/) and Nederland's National Institute for Public Health (RIVM/ https://www.rivm.nl/en/novel-coronavirus-covid-19/current-information-about-novel-coronavirus-covid-19).

From already mentioned resources we collected daily the following figures important for the model: the information about daily reported number of infected persons (INF/I), persons admitted to the hospital (H), people who deceased (Lost/F), people who recovered (R), active Healthcare workers (PS) and Healthcare workers infected (PSINFAC). We indirectly found the data about the number of people who are working even in this situation (TE) of the lockdown (beside healthcare workers, like drivers, those who deliver food, police etc.). We supposed that those who are in lockdown (Q) in the situation like this are close to the number of inhabitants, since self-isolation in both countries was mandatory. Latent persons (L) is people who are infected but asymptomatic or just not registered due to the lack of massive testing. L size can be inferred with some reliability from the biological research (16 times R in Spain in the first week of April, for example).

## 4. The Flow Network Model

This model is a combination of flow networks (FN) and the SIR model with some considerations from Markov chains. Figure 2 shows, as a starting point, the two networks: before (left) and after (right graph) the lockdown. In Spain a lockdown on 14-03-2020 and in the Netherlands partial one on 23-03-2020.

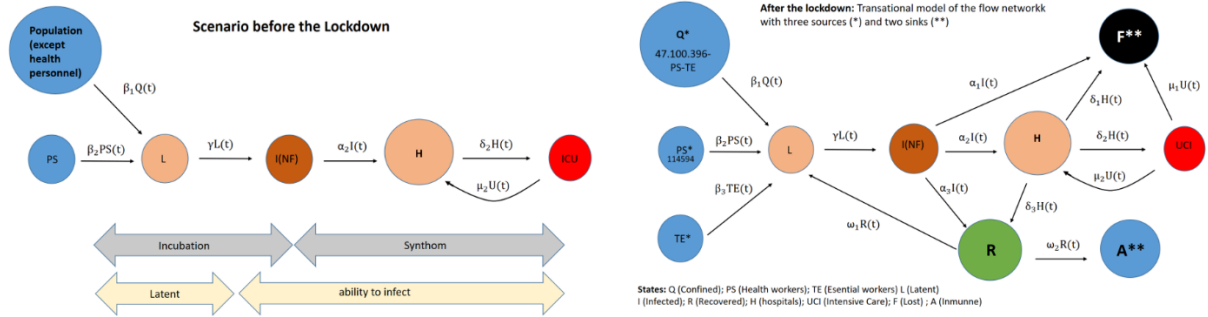

**Figure 2**. Flow Network Model SIR modified for the situation before (left graph) and after (the right graph) of the lockdown.

It can be seen on above graphical representation of the model that the nodes are symbolizing subpopulations of the number we begin with, the number of the whole population. Although it is an approximation of the reality (we simplify the problem in order to model it), the number from the beginning Q (those in lockdown/confined) and PS (active healthcare workers on the first line of primary care and in hospitals) are contributing to the next compartment L (Latent/Infected but not registered). On the version after the lockdown we added TE (the working people, who are actively in transfer during the lockdown). Then, after the exposure (or after the incubation which can be, according to WHO between 14 and 28 days) they gather in the compartment INF, which is Infected. Both L and INF can go directly to Recovered (R), but also those infected can die at home or die after admittance to the hospital (H). Those people who do recover (R) are coming from two different compartments, namely INF and H. Those who are treated in hospitals can develop life-threatening problems and be transferred to ICU; those from ICU can eventually become better after which they can be transferred to rehabilitation center or another department in hospital. The people who die (F, from Spanish Fallecidos) due to the consequences of COVID-19 are comprising of prior members of three different compartments: INF (those who die at home), H (those who die in the hospitals) and ICU (those who were on intensive care, probably on ventilators). The symbol A** represents those who, after being infected, recovered and develop the immunity on a corona virus (SARS-CoV-2). As we mentioned earlier it is still not clear from different sources of information, scientific literature included, whether the COVID-19 patients develop partial or full immunity, and whether re-infection is possible after recovery (we did not further specify this possible development in our model). Since we did not have reliable data about those who do have immunity on corona virus, and we cannot do this estimation, we considered in our model that two main end-nodes are R and F. The Objective functions we aimed to find in this research are the part of optimization, for which you need to introduce certain restrictions. For example, looking overall it is a question of in which relationship are Arrivals and Departures, and only if they are equal the system can be in equilibrium; but that is not possible in reality. We consider this work as a trial of making balancing this model due to the knowledge of certain transfers of possible importance for further management of resources. Hospitals and ICUs are in any system so far bottlenecks, since they have limited capacities. To manage to circumvent such a bottlenecks administrator(s) of a system can try to optimize the flow by tackling the certain transfers we managed to define.

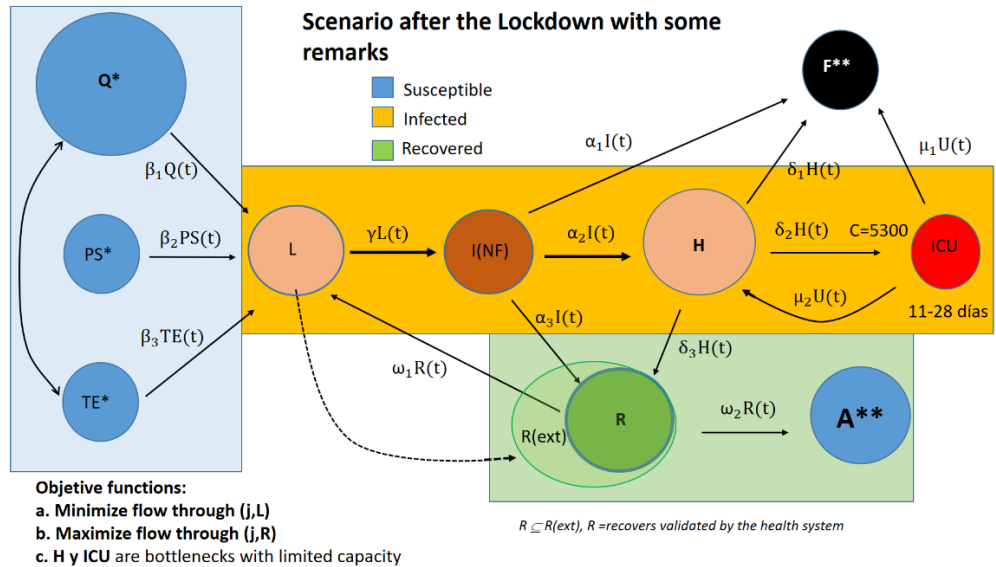

**Figure 3**. The scenario for the state of lockdown; SIR compartment model in combination with flow network and Markov chain additions.

Another question visible on our left panel (the model before the lockdown) is illustrated in the arrows in the bottom grey one -those still in incubatory period and those with symptoms of the disease, and yellowish one- showing those who are latent, or asymptomatic/without symptoms and those who are already infective (have capacity to infect others), so they can transfer the disease further in their contacts. On John Hopkins Website it can be seen for example, that more than 80.2% confirmed patients were actually with very mild symptoms, sometimes believing it was the seasonal flu. The number of asymptomatic people is still not known, because that would require massive testing of the whole populations or at least a big sample to estimate that number. Researchers from Germany, for example, found that almost half of total cases are originating from people without obvious symptoms. After a proper investigation and follow up they found that someone can be contagious two days before the development of symptoms (Dr. Drosten from Charite Berlin, Germany Epidemiological Coordinator).

The aim of all the measures applied at the moment is that the healthcare system and measures proposed would try to maximize the process/transfer/arrow leading from INF to R (and eventually to A**), and minimize the summary flow to F (Deceased).

Another representation of compartments leading us to the better understanding of the underlying dynamics can be seen on the Figure 3. As the Figure 3 shows, the SIR model fits perfectly within our Flow Network model. Per each arc (A,B) the flow is "number of persons that goes from A to B daily" which is a Poisson distribution where the parameter lambda is also variable.

Figure 4 is showing the set of equations behind the model. The first one is an invariant equation that make consistence of the model: Total population within the network is stable. The rest of the equations comes from the variation of the flow per each node in the same way than the differential equations used for SIR model.

$$Q+PS+TE+L+I+R+H+F+UCI+A=N, \ \forall t$$
$$Q(t+1)=Q(t) - \beta_1 \ Q(t) = (1- \beta_1) \ Q(t)$$
$$PS(t+1)=PS(t) - \beta_2 PS(t)$$
$$TE(t+1)=TE(t) - \beta_3 TE(t)$$
$$L(t+1)=L(t) + \beta_1 Q(t) + \beta_2 PS(t) + \beta_3 TE(t) - \gamma I(t)$$
$$I(t+1)=I(t) - \alpha_1 I(t) - \alpha_2 I(t) - \alpha_3 I(t) + \gamma L(t)$$
$$R(t+1)=R(t) + \alpha_3 I(t) + \delta_3 \ H(t) - \omega_1 R(t) - \omega_2 R(t)$$
$$F(t+1)=F(t) + \alpha_1 I(t) + \delta_1 \ H(t) + \mu_1 U(t)$$
$$H(t+1)=H(t) + \alpha_2 I(t) - \delta_1 \ H(t) - \delta_2 \ H(t) - \delta_3 \ H(t) + \mu_2 U(t)$$
$$U(t+1)=U(t) + \delta_2 \ H(t) - \mu_1 U(t) - \mu_2 U(t)$$
$$A^*(t+1)=A^*(t) + \omega_2 R(t)$$

**Figure 4**. Recurrent equations describing the process and trasfers we used in this dynamical model (except for the top one which is summary of the set).

Here we are giving the meaning of all the parameters used in recurrent equations, since they are the arrows illustrating the flow between compartments, including all the possible interactions we could include at the moment. Beta ($\beta$) is the average number of contacts per person per time (here we are choosing the time of update is one day), and decribes the transition from one compartment to another. We explored several scenarios to probe this. In case of our model, since we have three groups of susceptible peersons, there are three transition parpameters: probability of contagion/infection in isolation (for example members of household) $\beta_1$, probability of contagion for a person who is an active healthworker $\beta_2$, and probability of contagion of a person who works during the lockdown as a non-healthcare worker, $\beta_3$. The parameter Gamma ($\gamma$), the the probability of developing the illness once you get infected. L are people that got the virus but don't develop the illness (asymptomatic persons), INF are the people who after getting infected, are so evidently sick that they are easily registered. We did not mention the probability to change the state from latent to infective one (so one can transfer the disease) and on a final model there is not even an arrow between those two states. The parameters alpha are describing the probability of death: $\alpha_1$- the probability of dying at home (or in a nursing home), $\alpha_2$-the probability of going to a hospital, and $\alpha_3$- the probability of recovery at home. Parameter showing the probability of passing away in hospital (not in ICU) is described here as $\delta_1$; $\delta_2$ is probability of that one need a care in ICU, and $\delta_3$ is probability of recovering at hospital. Parameters signifying the probability of passing away in ICU ($\mu_1$) and $\mu_2$, the probability to leave the ICU and be transfered to another department in hospital (or another hospital for rehabilitation). Omega ($\omega_1$) signifies the probability to become infected for the second time (after the recovery) and ω2 signifies the probability of immunization (or of attaining the immunity to corona virus 2). In addition, we need to stress here that it is still not clear whether the presence of antibodies in the serum of a tested person is a certain support of total (or partial) immunity to SARS-CoV-2.

On the final model Figure 3 it can be seen that the L and INF/I nodes have been neighbors as they are logically connected by the flow of infection, affected by the viral load. According to the time windows, there is transit between the Q and TE nodes that are now Active Jobs outside the home. The model below does not study those immunized or infected in successive instances because they are residual or there is insufficient information. The model could be useful if applied to specific regions, such as a city or a town. The Spanish state is very diverse and this makes estimates very unreliable (our opinion is that Madrid as such a big affected city/a focus, is contributing to the public figures, but it is a question how that applies to other areas like Almeria or Alhambra for example). Thus the model had to be tested further with a fresh data. The model is very dynamic; the parameters require daily updates in order to obtain good quality results due to the constant introduction of political measures for the management of the pandemic. It is also a very vulnerable system, any misuse can lead to radical changes, such as neglect or misuse of prevention measures.

### Parameter estimation and balancing

$$\hat{\beta}_1(t) = \frac{\#contagiosQ(t)}{|Q(t-1)|} + \varepsilon_1(t)$$

$$\hat{\beta}_2(t) = \frac{\#contagiosPS(t)}{|PS(t-1)|} + \varepsilon_2(t)$$

$$\hat{\beta}_3(t) = \frac{\#contagiosTE(t)}{|TE(t-1)|} + \varepsilon_3(t)$$

$$\hat{\mu}_2(t) = \frac{|F(t) \cap U(t-1)|}{|U(t-1)|} + \varepsilon_{11}(t)$$

$$\hat{\mu}_3(t) = \frac{|H(t) \cap U(t-1)|}{|U(t-1)|} + \varepsilon_{12}(t)$$

$\varepsilon_i(t)$ factor of correction, can be set to zero for experimental purposes

$$\hat{\gamma}_1(t) = \frac{|I(t)|}{|L(t-1)|} + \varepsilon_4(t)$$

$$\hat{\alpha}_1(t) = \frac{|F(t) \cap I(t-1)|}{|I(t-1)|} + \varepsilon_5(t)$$

$$\hat{\alpha}_2(t) = \frac{|H(t)|}{|I(t-1)|} + \varepsilon_6(t)$$

$$\hat{\alpha}_3(t) = \frac{|R(t) \cap I(t-1)|}{|I(t-1)|} + \varepsilon_7(t)$$

$$\hat{\delta}_1(t) = \frac{|F(t) \cap H(t-1)|}{|H(t-1)|} + \varepsilon_8(t)$$

$$\hat{\delta}_2(t) = \frac{|U(t)|}{|H(t-1)|} + \varepsilon_9(t)$$

$$\hat{\delta}_3(t) = \frac{|R(t) \cap H(t-1)|}{|H(t-1)|} + \varepsilon_{10}(t)$$

**Figure 5**. Estimation of parameters used must be done regularly for balancing the model.

Based on above described model, a code in R is written and fed by the data from official sources mentioned in the Data section, which were updated daily. This yielded an evolution of the states cumulatively from early March to April 29th 2020 for Spain, and as a comparison similar output graphic representation for the Netherlands (although with some missing data, as we could not retrieve the number of recovered persons from RIVM website, we only combined it with some updates from JHU). We need to stress once more that our model is a generalization based on the data from the whole country, but we are confident that it would work even better for a smaller region or data with easily stamped data.

## 5. Results

Our results comprise of two parts: the first one is the evolution of the spread of disease in Spain from 9-03-2020 to 29-04-2020, with a special focus on a novel trend obtained in the third week of April, and the cumulative evolution of a spread in the Netherlands, for the same period. Note that the proposed lockdowns were ten days apart. There are some specific remarks we can infer from those results:

- The model indicates that the probability of contagion is higher for active sanitary workers ($\beta_2 \cong 8.9*10^{-3}$). Logically, those in isolation are the least vulnerable ($\beta_1 \cong 0.07*10^{-3}$). Estimated beta parameter for regular workers (under the recommended protection) is bounded as $\beta_3 <= 3*10^{-3}$ (a proper estimation is not possible due to lack of data).

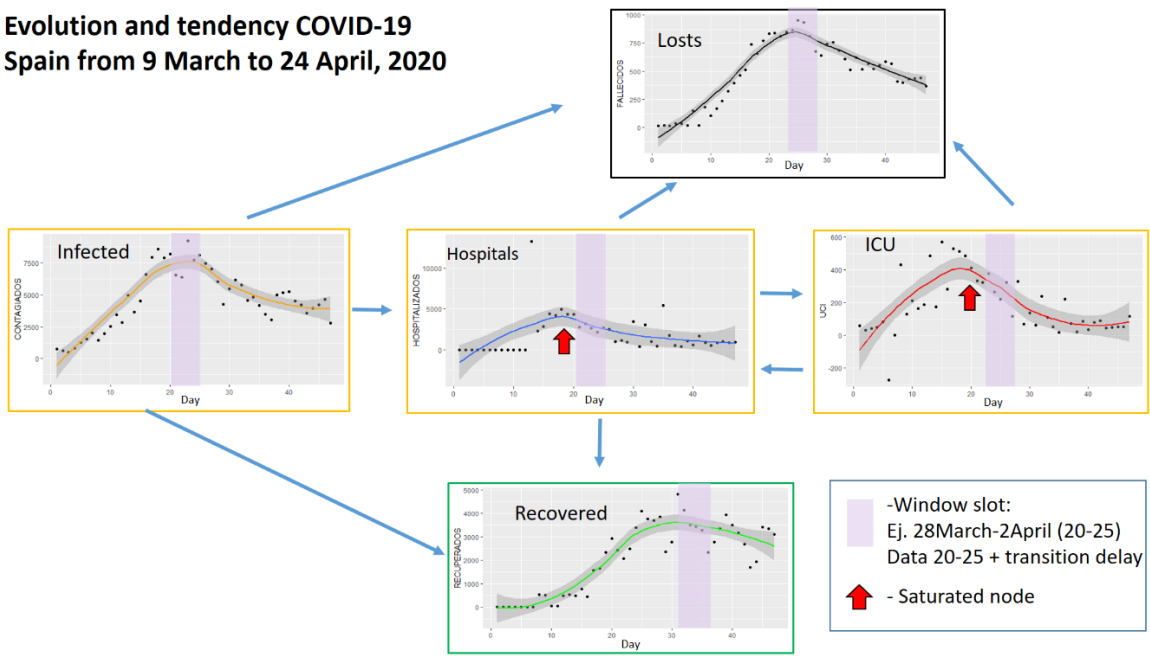

**Figure 6.** Evolution and tendencies of COVID-19 for Spain (from 9 March to 26 April 2020)

- As from Figure 3 it can be seen that for further optimization administrators have to minimize all the inflows to L and to maximize all the transfers to R. Hospitals and ICUs are bottlenecks due to a limited capacity (illustrated with red arrows indicating the saturation of those nodes).

- Within the flow network there are two types of optimization (1-2): minimization of the input flow $OF1$ (flow to vertices INF and F –Lost) and maximization of the output flow $OF2$ (flow to R vertex),

$$OF1 = min \sum_{x \in V} f_{x,INF} + \sum_{x \in V} f_{x,F} \qquad (1)$$

$$OF2 = max \sum_{x \in V} f_{x,R} \qquad (2)$$

+

with these aims, a "danger index" (DI) is calculated as (3) every day (∀t),

$$DI(t) = \sum_{x \in V} f_{x,INF}(t) + \sum_{x \in V} f_{x,F}(t) - \sum_{x \in V} f_{x,R}(t) \quad (3)$$

Figures 7 and 8 shows the calculations for the period 9 March to 22 April (Figure 7) and to 26 and 30 April (Figure 8). The figures show as well the regression curve associated with the point cloud. Substantial differences between the curves are observed, indicating the sensitivity of the data.

- As a conclusion of this study, we could say that these data indicate a *danger of rebound in the pandemic*. As the first 'peak' of infections in Spain was behind us (reached on march 26th), people gradually start going out for work, but it is again giving rise to the number of newly infected ones and it can mean that another peak of spread is following.

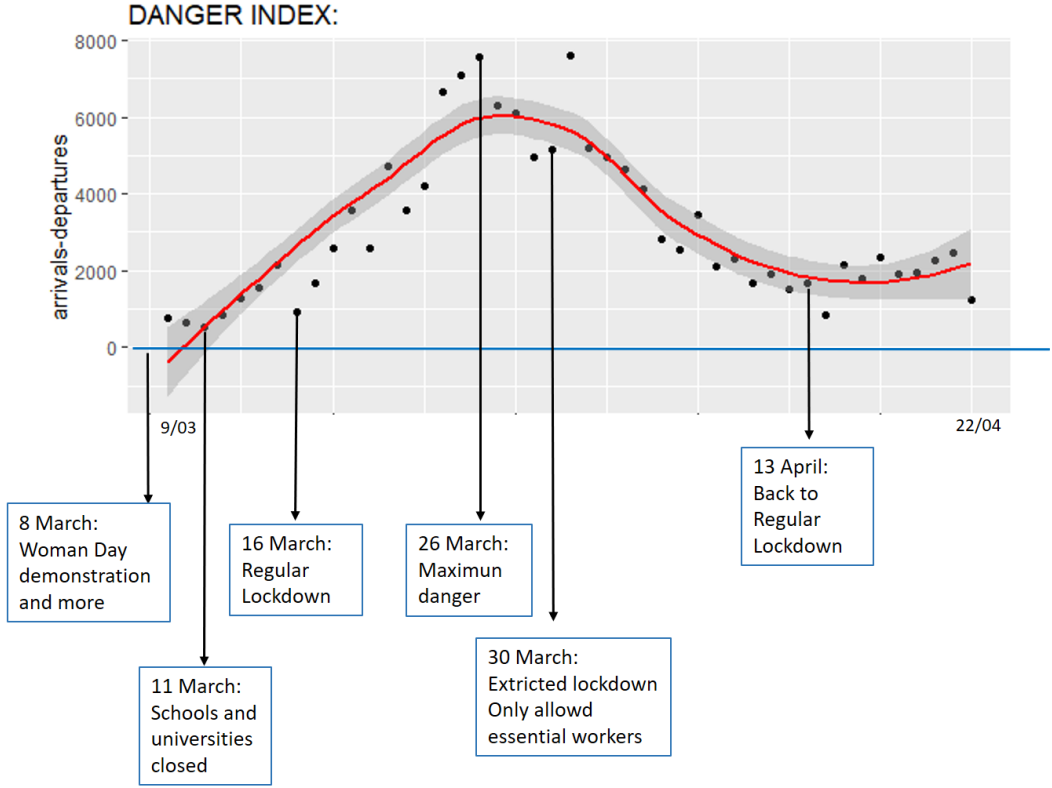

**Figure 7.** The danger index on March 22th.

Figure 7 shows the "danger index", which is a formula that computes the difference between the arrivals and departures in the network. This difference is desirable to be negative. Linear regression curve indicates an increasing tendency. It can be seen that the 'peak' of spread dynamics was on March 26th. But if we observe the difference between the inflow and output from this network, we can see at the far right part of the graph, that another rise of infected persons is starting again. Figure 8 shows same approach 4 days after. We can observe how the tendency can change after a few new input data. In this figure there are also remarks to some important dates to understand the evolution of the sanitary crisis in Spain. 26th April changes totally the tendency of the red line, which is here touching the blue line (blue is symbolizing the confidence interval).

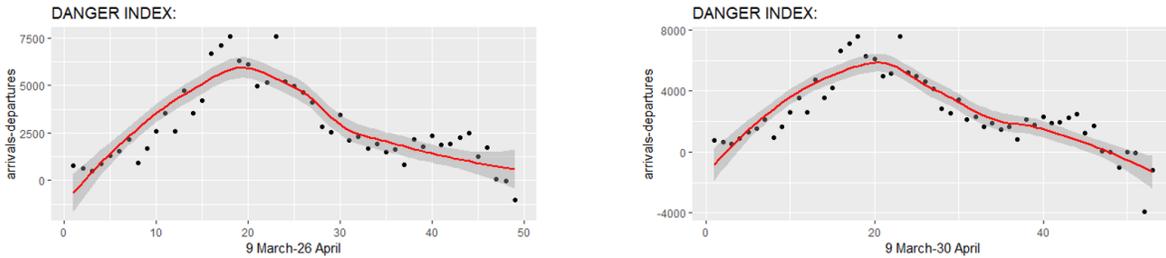

**Figure 8**. The danger indicator with information on 26 April. The curve can start rising again.

In Comparison to Spain, Netherlands are lagging in time with quite similar dynamics illustrated in next figure.

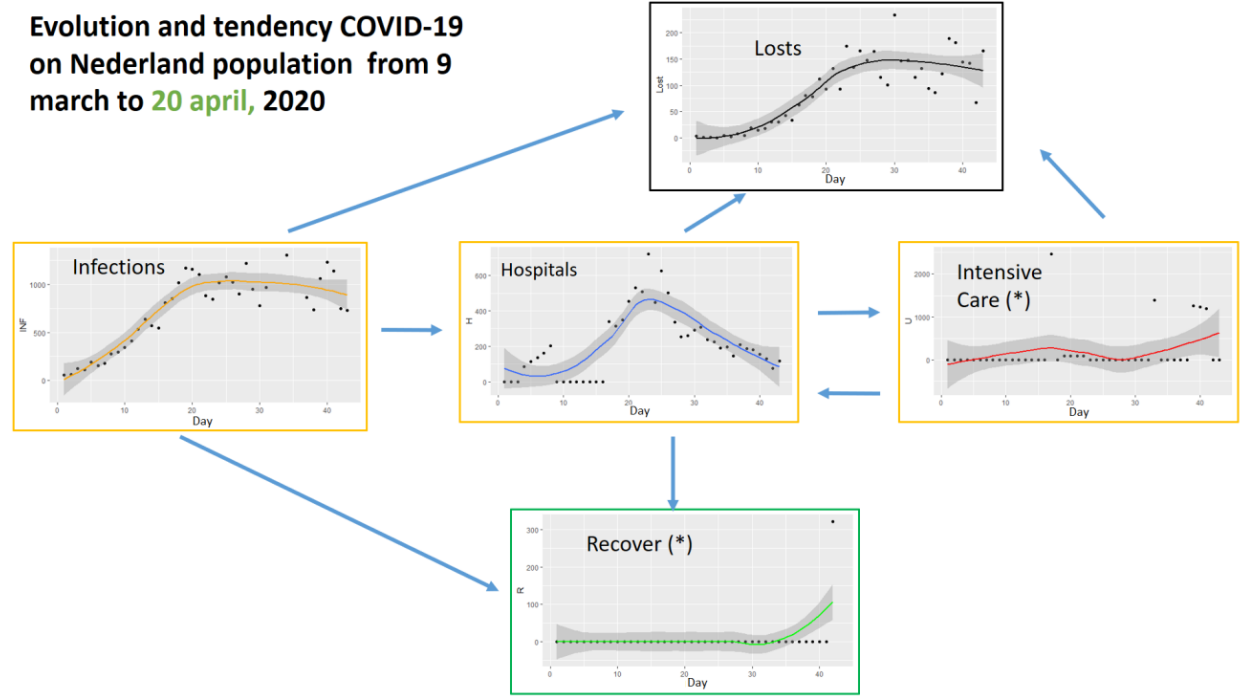

**Figure 9**. The evolution and observed trends in the Netherlands. Due to some mising data, we were not able to estimate the number of recovered persons, but it seems, as also stated on RVO website that the proposed measures of social distancing, self-isolation and working from home are giving the results.

The problem we faced is that the data reported by RIVM in the Netherlands are limited to the total number of registered infected persons, (and new daily registered cases), number of people admitted to the hospitals (with a daily number of new admissions) and the number of registered deaths (with the disclaimer that the numbers may vary since nor the hospital admissions not the deceased are accurate due to specific procedures causing aggregation of the data for several days).

The number of health professionals infected we got indirectly from another study published by the group in Erasmus Hospital in Rotterdam, and the number of recovered is totally absent. We found some data on JHU, but the total number of 250 recovered people when the total number of cases was already above 30 000 is probably unreliable. We tried to infer some data from those which we have, with the help of La Place's rule, but sometimes we could not do that either. Another information which is missing is the number of people in ICUs; again we found just sporadic data presented on other news during the pandemic, like 1200 persons in total, according to governmental media. Also, from other sources of information we learned that a portion of patients from Netherlands was treated in neighboring Germany, due to their better preparedness. Some cases were, for example, transferred from Rotterdam to Amsterdam, and all in all, we are aware that the accurate figures will be known only after the whole crisis would be over. So we decided to focus on the part of the model we can infer about, and the final nodes which were R and F   (we already stated that we decided not to elucidate about A** since the mechanism of attaining immunity is still not known entirely).

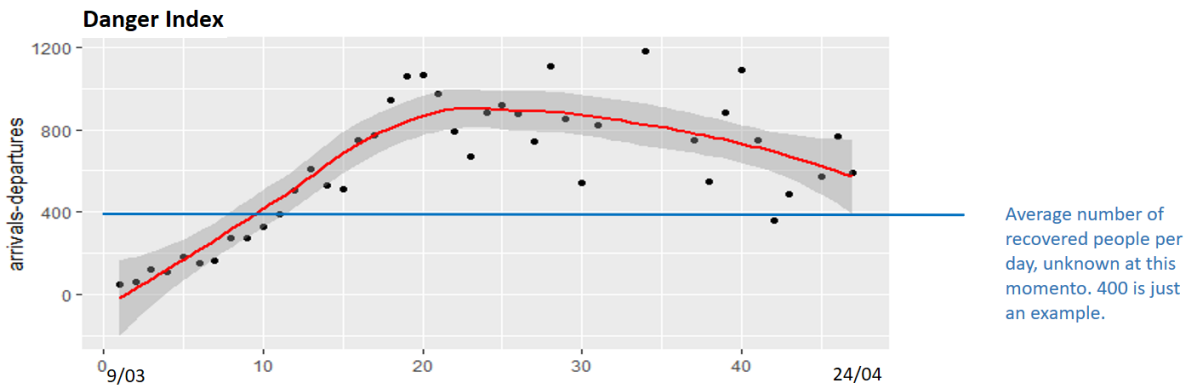

**Figure 10**. Danger index for Netherlands. As the number of recoverd people is not published, a blue line is used as a reference to make the decision. In de picture it is set an average of 400 persons per day recoverd, as an example.

Again, when we take into account all the unknowns, we suppose that the rebound is possible, which could be only seen after the gradual relaxation of special measures imposed to all the population. On the last day we entered the data to the model, the numbers of admitted patients in hospitals, the registered deceased and the newly registered cases dropped significantly (88, 48,171). But according to already mentioned study performed with the support of Sanquin (Royters), only 3% of tested population developed antibodies, which (translated to the whole population) is pointing out at the level of potential immunity much lower than desired herd-immunity which was so overrated term in a public discourse.

### 6. Discussion and conclusions for future work

Our results show, by employing our Dynamical model fed with publicly available data, the SARS-CoV-2 (which cause the disease COVID-19) spread evolution and trends in Spain and in the Nederland cumulatively from 9 March to 29 April 2020. It is clear that in both countries the recommended measures of social distancing and closing the public gatherings are giving the results; the 'peak' of new infections, admissions to hospitals and deaths already happened in both countries. We think that our dynamical model based on compartment SIR epidemiological model with modifications from Markov chain and network flow are giving the tool to those who want to optimize the management of resources especially in other than central areas. It is of importance since our results are also showing that another wave of infections is possible and that we would probably have another peak(s).

Our model is giving indications of future increase in infections which is in line with other researcher's findings that only a fraction of population (6.4%) in any European country will have the antibodies necessary for later immunity on COVID-19 (Moran et al., 2020). It is still not clear whether those who recover after the infection are attaining partial or total immunity for COVID-19, since there are reports of sporadic re-infections in Asia. A study of Dutch blood donors has found that around 3% have developed antibodies against the new coronavirus; which change the insight in how many people were being sick below the radar of detection (Reuters). On a population of above 17 million, it is much more than the publicly available records (giving the retrograde conclusion that some half of a million were sick in the past). In the beginning of March an analysis of waste water around Amsterdam showed the presence of corona virus, illustrating the extent of spread of infection, which is probably larger than the detected cases (which is not strange since only around 17000 tests were performed in the Nederland until the conclusion of this study). Our results are implying that active health workers have increased probability of contagion and that only those in isolation are safe, but that is defensive measure. Since the first vaccines can be expected in around a year (for safety reasons) the question here is how we are going to live, since the majority of population is not immune to the virus infection? Many will agree that life will have to change after this pandemic, significantly.

Another important issue in this crisis is the role that asymptomatic persons are having. In Germany, it is found (an interview with the country's medical coordinator Dr. Christian Drosten with Guardian) that almost half of all infections was coming from contagious but asymptomatic people. At least there were without symptoms in the time of transmission; another follows up showed that they became contagious two days before developing the symptoms. Dr. Drosten Is pointing out that the tracking mobile systems are racing with the time, since the dynamic is very complicated when those latent people are in the game. There is also another effect Dr. Drosten mentions, which is the irrational effect of their successful strategy; Germany has the lowest number of deceased in regard with the total number of registered infections which is giving the false impression that the problem is exaggerated and that those who are strategists of epidemiological health workforce are against economy. Every country has its own flaw our, like Netherlands showing overrated optimism in the beginning of the crisis, demonstrating false impression of safety versus small risk perceived.

We are aware that our study has some limitations. We think that among limitations and restrictions (which should be taken in consideration) of our model has the following:

• Model assumes a constant N population (no births or immigration, as any SIR model)

- The population is not homogeneous, this study would be probably more useful for a small region (although, Moran et al., 2020 showed that the small city region model is less effective than the dynamical one)
- The parameters are estimated from the official data (which are sometimes aggregated)
- Q represents the quarantined and healthy node, i.e. one person may be incubating the virus and being in the quarantine, but that person is placed in the latent node L
- The strength of the infection and rate of recovery determine some transfers
- No pharmacological intervention (vaccines, drugs) exist at the moment
- Transmission is by random contact (all persons have the same probability of exposure) - after the state of Lockdown and within the corresponding population sector)
- Infection capacity per infected person (2 to 3 people), some reports indicate 2.5 or somewhat less
- The initial values on the population Q(0), PS(0) and TE(0) are also estimates from the official data.

We believe that our model can be even better applied to smaller regions and even cities with their collected data, since the publicly available data that we used is probably affected greatly with the big cities like Madrid and Amsterdam, for example, and the dynamics can be somewhat different in another province. Therefore, we would like other researchers to test further this model.

In conclusion, we can say that our results show that the dynamical trend in Nederland seems to follow a very similar pattern in the Spain, it is just lagging behind for less than two weeks. Due to the lack of R data on the Nederland side we can only estimate it from our present data. As a conclusion of this study, we could say that these data indicate a danger of rebound in the pandemic. Since the data show that Spain already reached the 'peak' of spread, it could mean that other 'peaks' are to follow, which can inform further preventive decisions from authorities. That can be concluded also with the comparison with the most similar pandemic of Spanish fly in 2018; it has three peaks in total, and the policy makers affected the curve of spread greatly.

In a documentary about pandemics from Netflix serial 'Explained', emitted before the onset of SARS-CoV- 2 in Europe, a sentence from an expert epidemiologist from WHO stays a long time with us: 'Mother nature is the ultimate bioterrorist'. The narrative was (and now we all know too well) that the question is not 'will the pandemic hit us' but 'when'. We hope that we can at least, learn from the previous experience and make better optimization of our health care systems in next period. Because, the virus is here to stay. And we have to find a way to survive it. The first step to that goal is to understand every little detail in the dynamics of its spread. If almost every state in the world is doing the simulation of future elections or invest heavily in warfare, then the usage of precisely the same techniques and resources for fighting the virus should be possible.

***Additional***: *More information and graphics can be download from*
https://github.com/vlopezlo/Covid_19/. *To those who would like to perform the same analysis on their regional data, please, ask for the access to the R code and datasets (vlopezlo@ucm.es).*